\pdfoutput=1

\documentclass[12pt]{article}
\usepackage[
top = 2.5cm, 
bottom = 2.5cm, 
left = 2.5cm,  
right = 2.5cm]{geometry}

\usepackage{latexsym} 
\usepackage{verbatim}
\usepackage{tikz}
\usetikzlibrary{matrix}
\usepackage{color}
\usepackage{graphicx,amssymb,amsfonts,amsmath,amssymb,amscd,amstext, mathrsfs}
\usepackage{graphicx} 
\usepackage{bbm}
\usepackage{dsfont}
\usepackage{xcolor}
\usepackage[colorlinks=true, citecolor=black, linkcolor=black, allcolors=black]{hyperref}   
\usepackage{amsmath}
\usepackage{empheq}
\usepackage{hyperref}

\newlength\dlf

\def\be{\begin{eqnarray}}
\def\ee{\end{eqnarray}}
\def \bea {\begin{equation}}
\def \eea {\end{equation}}

\def \nn {\nonumber}

\def \si{\sigma}

\def \d {\partial}

\def \rr {\raise.35ex\hbox{\small $\prime$}\kern-.17em{\mbox{\large $\imath$}}}

\def \dels {\partial\kern-.5em / \kern.5em}
\def \As {{A\kern-.5em / \kern.5em}}
\def \Ds {D\kern-.7em / \kern.5em}

\def\frac#1#2{{#1\over #2}}

\def \iffa {\iffalse} 

\def \ed  {\end{document}}
 
\def \p {\partial}

\def\nn{\nonumber}

\def \ed  {\end{document}}

\DeclareFontFamily{U}{mathx}{\hyphenchar\font45}
\DeclareFontShape{U}{mathx}{m}{n}{<-> mathx10}{}
\DeclareSymbolFont{mathx}{U}{mathx}{m}{n}
\DeclareMathAccent{\widebar}{0}{mathx}{"73}
\def\zbar{{\widebar z}}

\DeclareFontShape{OT1}{cmr}{mx}{n}{<->cmr10}{}
\newcommand{\titlefont}{\fontseries{mx}\selectfont}

\def \si{\sigma}

\def \I{{\cal I}}

\usepackage{cite}

\begin{document}
\begin{titlepage}

\begin{flushright} 
\end{flushright}

\begin{center} 
\vspace{2.5cm}  

{\fontsize{18.5pt}{0pt}{\titlefont{Approximate Symmetries in\\
\vspace{0.6cm}  
$d=4$ CFTs with an Einstein Gravity Dual}}
\vspace{1.5cm}  
\\
 {\fontsize{12pt}{0pt}{\titlefont{Kuo-Wei Huang}}}}   
\\ 
\vspace{0.6cm} 
{\it{Department of Physics, Boston University,\\ Boston, MA 02215, USA
}}\\
\end{center}
\vspace{0.3cm} 

\begin{center} 
{\titlefont \bf{Abstract}}
\end{center}
\vspace{-0.3cm} 

By applying the stress-tensor-scalar operator product expansion (OPE) twice, we search for algebraic structures in $d=4$ conformal field theories (CFTs) with a pure Einstein gravity dual. We find that a rescaled mode operator defined by an integral of the stress tensor $T^{++}$ on a $d=2$ plane satisfies a Virasoro-like algebra when the dimension of the scalar is large. The structure is enhanced to include a  Kac-Moody-type algebra if we incorporate the $T^{--}$ component. In our scheme, the central terms are finite. It remains challenging to directly compute the stress-tensor sector of $d=4$ scalar four-point functions at large central charge, which, based on holography and bootstrap methods, were recently shown to have a Virasoro/${\cal W}$-algebra vacuum block-like structure.

\end{titlepage}

\addtolength{\parskip}{0.6 ex}
\jot=2 ex

\subsection*{1. Introduction}

The aesthetic appeal of symmetry has been a guide for physicists. In the context of Conformal Field Theory (CFT), the Virasoro symmetry provides a powerful constraint that leads to a partial classification of two-dimensional CFTs \cite{BELAVIN1984333}. The notion of symmetry continues to evolve. Many symmetry algebra-related techniques used in understanding $d=2$ CFTs, however, do not find applications in higher dimensions. One may wonder, by imposing certain physical conditions, if a similar algebraic structure emerges in a class of $d>2$ CFTs. Effective symmetries will allow one to improve the bootstrap program on CFTs in higher dimensions \cite{Poland:2018epd}.

There are some examples: supersymmetric CFTs in $d=3$, $4$, and $6$, with at least 8 real supercharges, have a subsector described by a chiral algebra \cite{Beem:2013sza, Bobev:2020vhe}. Moreover, the maximally supersymmetric $d=6$ CFTs have an ${\cal W}_N$ algebra \cite{Beem:2014kka}.  An earlier work \cite{Banados:1999tw} (see also \cite{Brecher:2000pa}) discussed the possibility of extending the Brown-Henneaux symmetry \cite{Brown:1986nw} to higher dimensions using traveling waves on the anti-de Sitter (AdS) background.

From gauge/gravity duality's viewpoint, the simplest CFTs are those which have a pure gravity dual description.  They are believed to be strongly coupled, large $N$ CFTs with an infinitely large gap to the lightest higher-spin single-trace primary \cite{Heemskerk:2009pn, Camanho:2014apa, Afkhami-Jeddi:2016ntf, Meltzer:2017rtf, Belin:2019mnx}. In this work, we attempt to search for approximate Virasoro-type structures in $d=4$ CFTs with an Einstein gravity dual.  By approximate, we mean that algebraic structures emerge under certain physical conditions. There have been several recent works devoted in this direction \cite{ Huang:2019fog, Huang:2020ycs, Belin:2020lsr, Besken:2020snx, Huang:2021hye, Korchemsky:2021htm}; the approach used in this paper will be different. 

Is there a concrete result that motivates us to search for a Virasoro-type algebraic structure in $d=4$ CFTs? One motivation comes from the recent progress of holographic and conformal bootstrap computations of a scalar four-point function in $d>2$ CFTs with a large central charge \cite{Fitzpatrick:2019zqz, Li:2019tpf, Kulaxizi:2019tkd, Fitzpatrick:2019efk, Karlsson:2019dbd, Li:2019zba, Karlsson:2019txu,  Karlsson:2020ghx, Li:2020dqm, Parnachev:2020fna, Fitzpatrick:2020yjb,  Parnachev:2020zbr,  Karlsson:2021duj, Rodriguez-Gomez:2021pfh, Rodriguez-Gomez:2021mkk, Krishna:2021fus, Karlsson:2021mgg}. Using AdS effective field theory, the operator product expansion (OPE) coefficients corresponding to multi-stress-tensor exchanges were computed for $d=4$ CFTs dual to semiclassical Einstein gravity with a minimally coupled scalar \cite{Fitzpatrick:2019zqz}. In $d=2$, these OPE coefficients are fixed by the Virasoro symmetry. To extract the OPE coefficients in $d>2$, the authors of \cite{Fitzpatrick:2019zqz} considered a two-point function of the light probe scalar in a black-hole background created by heavy operators. The bulk effective action reads
\begin{align}
S_{\rm {EFT}} \sim C_T \int d^{5}x \sqrt{g} \big(R+\Lambda \big)+ {1\over 2} \int d^{5}x \sqrt{g}  \big((\partial \phi)^2 +  m^2 \phi^2  \big) 
\end{align}\vspace{1mm}where $C_T \sim G^{-1}_N$ is the central charge. The light scalar $\phi$ has a finite mass at a large $C_T$.  In this setup, one can first solve for the metric while ignoring the probe scalar, then solve for the scalar two-point function in the black-hole background. In the boundary limit, the two-point function can be identified as a heavy-light four-point correlator: 
\begin{align}
 \langle  {\cal O}_H  {\cal O}_H {\cal O}_L {\cal O}_L \rangle ~~~~~ {\rm with} ~~ C_T  \to \infty~,~{\Delta_H \over C_T}~ {\rm fixed}
\end{align} 
where ${\cal O}_L$ is the boundary dual of $\phi$. The higher-dimensional analogue of the $d=2$ Virasoro vacuum block is the contribution to the four-point function in the channel ${\cal O}_H {\cal O}_H\to {\cal O}_L{\cal O}_L$ from $d>2$ multi-stress-tensor exchanges. See \cite{Hartman:2013mia, Fitzpatrick:2014vua, Fitzpatrick:2015qma, Fitzpatrick:2015zha, Perlmutter:2015iya} for the $d=2$ case.\footnote{It turns out that, in a  lightcone limit, the $d>2$ scalar correlators are not affected by higher-curvature corrections in the gravitational action if one assumes minimal coupling.  Certain non-minimally coupled  interactions can modify the near-lightcone correlators \cite{Fitzpatrick:2020yjb}.} In \cite{Kulaxizi:2019tkd, Karlsson:2019dbd, Karlsson:2020ghx}, it was observed that the $d>2$ OPE coefficients computed via AdS/CFT \cite{Fitzpatrick:2019zqz} satisfy a certain close form. The stress-tensor sector of the near-lightcone correlator in even-dimensional CFTs with an Einstein gravity dual, when expressed as a sum of $n$ stress-tensor exchanges, has the following Virasoro/${\cal W}$-algebra vacuum block-like structure:   
\vspace{1mm}
{\small
\begin{align}
\label{PANA}
{\lim_{\zbar \to 0}} \langle  {\cal O}_H(\infty)   {\cal O}_H(1) {\cal O}_L(z, \zbar) {\cal O}_L(0)\rangle|_{\bf \Sigma} &=  {1\over (z \zbar)^{\Delta_{L}} }  \sum_n  \sum_{\{i_p\}}   a^{(d)}_{i_1 \cdots i_n} f_{i_1}(z) \cdots f_{i_n}(z) \big({\Delta_H\over C_T}\big)^n {\zbar^{{n\over 2}{(d-2)}}}   \\
f_{a}(z)&=z^{a}~{}_{2}F_{1}(a,a,2a;z)  \\ 
\sum_{p=1}^n i_p &=  {n\over 2}({d+2})   
\end{align}}\vspace{1mm}The $d-2$ transverse coordinates are set to zero via a conformal transformation. We denote ${\bf \Sigma}$ as a two-dimensional Euclidean plane. The coefficients $a^{(d)}$ depend only on the spacetime dimension and the dimension of the light scalar $\Delta_L\equiv \Delta$.  The Virasoro algebra determines this correlator when $d=2$ (i.e. the Virasoro vacuum block).  The $d>2$ correlator structure \eqref{PANA} is consistent with all known OPE coefficients obtained by AdS/CFT \cite{Fitzpatrick:2019zqz}. Additional consistency checks based on the conformal bootstrap were performed in \cite{Karlsson:2019dbd, Karlsson:2020ghx}.

It is interesting that the $d=2$ Virasoro block structure extends to $d>2$ in the lightcone limit (i.e. $\zbar \to 0$).\footnote{The single stress-tensor exchange is fixed by the Ward identity.  Schematically, for two stress-tensor exchanges, the $d=2$ Virasoro vacuum block has the form $\sim f_2 f_2+f_1 f_3$. 
The $d=4$ near-lightcone correlator has the form $\sim f_3 f_3+ f_2 f_4+ f_1 f_5$.  A  similar pattern extends to any even dimensions.  If one defines the lightcone limit as $z \to 0$ with $\zbar$ fixed, the correlator is the same as \eqref{PANA} with $z, \zbar$ swapped.} The $d>2$ Virasoro block-like structure does not guarantee that it has a Virasoro-type symmetry origin, but we would like to make progress towards a better understanding of such a $d>2$ CFT correlator. 

The $T{\cal O}$ OPE and $TT$ OPE are the underlying structures of the correlator \eqref{PANA}. We aim to search for approximate symmetries when stress tensors are inserted in a scalar correlator.  It appears that an immediate difficulty is to identify the correct limiting procedure.  

In a recent work \cite{Huang:2021hye}, a Virasoro-like algebra was obtained via  the $d=4$ $TT$ OPE. The author defined a mode operator by integrating the transverse coordinates of the stress tensor. Such a ``purely" $d=2$ structure in $d=4$ theories does not seem to fully capture the $d=4$ near-lightcone correlator structure.  In the same work, the author observed that, by setting transverse coordinates directly to zero, one can reproduce the $d=4$ single stress-tensor exchange result via a mode summation. Another motivation of this paper is thus to study more about the approach in which one sets transverse coordinates to zero in the OPE. Instead of using the $TT$ OPE, we will use the $T{\cal O}$ OPE.

While the correlator \eqref{PANA} is well-defined, the $d>2$ stress-tensor commutators have divergent central terms. Therefore, a proper renormalization is required, presumably, to link the stress-tensor commutators to the scalar correlator. In this work, we introduce renormalized mode operators that allow us to remove the divergence. The main observation is that these mode operators defined by integrals of the stress tensors $T^{++}$ and $T^{--}$ satisfy a Virasoro-Kac-Moody-like algebra, when the scalars have a large dimension.  The novelty of our finding is that the emergence of $d=4$ stress-tensor-mode algebras depends on the properties/limits of the scalars. Crossing the logical chasm between these commutators and the near-lightcone correlator remains a challenging task, but we hope that this work can be useful towards efficiently solving the stress-tensor sector of $d=4$ CFTs with a pure gravity dual.

This paper is organized as follows:  in Section 2, we review the story in two dimensions as a warm-up example.  In Section 3, we discuss the $d=4$ $T{\cal O}$ OPE, central terms, mode operators, and commutators.  We close with discussion in Section 4.  Details of the $T{\cal O}$ OPE are listed in Appendix.

\subsection*{2. $d=2$ CFT} 

The $d=2$ holomorphic OPE between the stress tensor and a chiral scalar of weigh $h$ is 
\begin{align}
T(z){\cal O}(w)= \Big({h \over (z-w)^2}+  {1 \over z-w} \partial_w  \Big) {\cal O}(w) + \rm {regular} \ .
\end{align} 
The Virasoro modes $L_m$ can be expressed as a differential operator:
\begin{align}
\label{2dD}
[L_m, {\cal O}(w)]=  \oint_{{\cal C} (w)} {dz\over 2 \pi i}  z^{m+1}  T(z){\cal O}(w) &= \big( h (m+1) w^m+w^{m+1} \partial_w  \big){\cal O}(w) = D^{(w)}_m {\cal O}(w)  \ .
\end{align}
We are interested in the commutator inserted in a scalar correlator:
\begin{align}
\label{8}
\langle [L_m, L_n] {\cal O}(w) {\cal O}(0)\rangle \equiv \langle L_m L_n {\cal O}(w) {\cal O}(0)\rangle - \langle L_n L_m {\cal O}(w) {\cal O}(0)\rangle \ .
\end{align}
With two stress tensors, $\langle L_m L_n {\cal O}(w) {\cal O}(0)\rangle$ contains both the $T{\cal O}$ and $TT$ OPE contributions. 
Let us write
\begin{align}
\label{9}
\langle L_m L_n {\cal O}(w) {\cal O}(0)\rangle=  \langle L_m L_n {\cal O}(w) {\cal O}(0)\rangle|_{T{\cal O}}+  \langle [L_m, L_n] {\cal O}(w) {\cal O}(0)\rangle \ .
\end{align}
From \eqref{8} and \eqref{9}, we obtain
\begin{align}
\label{formula}
\langle [L_m, L_n] {\cal O}(w) {\cal O}(0)\rangle|_{T{\cal O}}= - \langle [L_m, L_n] {\cal O}(w) {\cal O}(0)\rangle  \ , 
\end{align}
where $\langle [L_m, L_n] {\cal O}(w) {\cal O}(0)\rangle|_{T{\cal O}}\equiv\langle L_m L_n {\cal O}(w) {\cal O}(0)\rangle|_{T{\cal O}}- ( m \leftrightarrow n)$ only includes the $T{\cal O}$ OPE contribution.  
We will use this relation to deduce algebraic structures without relying on the full $TT$ OPE in both $d=2$ and $d=4$.   
The central term  has to be added separately using the stress-tensor two-point function. 
This approach can be generalized to include two different mode operators.  

Taking $\langle {\cal O}(w) {\cal O}(0)\rangle= w^{-2h}$, we have\footnote{One can consider a general form $\langle {\cal O}(w) {\cal O}(z)\rangle= (w-z)^{-2h}$, but it does  not affect the final algebra.} 
\begin{align}
 \langle[D^{(w)}_m, D^{(w)}_n]  {\cal O}(w) {\cal O}(0)\rangle|_{T{\cal O}}&= - h (m - n) (m + n-1) w^{m+n-2h} \nn\\
&= - (m-n) \langle D^{(w)}_{m+n}  {\cal O}(w) {\cal O}(0)\rangle   \ . 
\end{align}
This corresponds to the  $d=2$ Witt algebra, $[L_m, L_n]=   (m-n)  L_{m+n}$.    

One can skip the step of obtaining a differential operator for $L_m$ and calculate the commutator  by the following operation:  
\vspace{1mm}
{\small
\begin{align}
\langle [L_m, L_n] {\cal O}(w) {\cal O}(0)\rangle= - \Big( \oint_{{\cal C}(w)} {dz_1\over 2 \pi i}  \oint_{{\cal C}(w)} {dz_2\over 2 \pi i} z_1^{m+1} z_2^{n+1} \langle T(z_1)T(z_2) {\cal O}(w) {\cal O}(0) \rangle|_{T{\cal O}~}  -  \big( m \leftrightarrow n \big) \Big)
\end{align}}\vspace{1mm}where  $\langle \cdots \rangle|_{T{\cal O}}$ includes only the $T{\cal O}$ OPE contribution.   After adding the central term defined by the stress-tensor two-point function,
\begin{align} 
\langle  T(z_1) T(z_2)\rangle= {c\over 2 (z_1-z_2)^4} ~~  \Rightarrow  ~~\langle  [L_m, L_n] \rangle  = {c\over 12} m (m^2-1) \delta_{m+n,0} \ ,
\end{align}
the Virasoro algebra is 
\begin{align}
[L_m, L_n]=   (m-n)  L_{m+n}+ {c\over 12} m (m^2-1) \delta_{m+n,0}    \ .
 \end{align}
The structure is similar for the anti-holomorphic part.  Certainly, this is a well-known result, but the point here is simply that one can deduce the algebra by applying the $T{\cal O}$ OPE twice. We will adopt this approach in $d=4$.

\subsection*{3. $d=4$ CFT}
\subsubsection*{\it{3.1. $T{\cal O}$ OPE}}

We are interested in a class of $d=4$ CFTs with a pure gravity dual. In this case, even for higher-point functions, only ${\cal O}$ appears in the $T{\cal O}$ OPE.  See \cite{Belin:2019mnx, Meltzer:2017rtf} for recent discussions.     

One can derive the $T{\cal O}$ OPE  by matching the most general tensors built out of $\delta^{\mu\nu}$ and $s^\mu= x^{\mu}_1-x^{\mu}_2$ to the short-distance limit (i.e. $x_1 \to x_2$) of the three-point function, 
\vspace{1mm}
{\small
\begin{align}
\label{TOO}
\langle T^{\mu\nu}(x_1){\cal O}(x_2){\cal O}(x_3)\rangle&= \frac{a}{x_{12}^4 x_{13}^4x_{23}^{2\Delta-4} } \big( \frac{X^{\mu}X^{\nu}}{X^2}-\frac{\delta^{\mu\nu}}{4} \big) \ , ~~ a= - \frac{2\Delta}{3\pi^2} \ , \\
 X^{\mu} &= \frac{x_{12}^{\mu}}{x_{12}^2}-\frac{x_{13}^{\mu}}{x_{13}^2}  \ .
\end{align}
}\vspace{1mm}The first two terms in the OPE were obtained long ago \cite{CARDY1987355, Osborn:1993cr}; several higher-order terms were computed recently \cite{Belin:2019mnx}. We have simplified their expressions and extended the computation to include one higher-order term, denoted as $F$. The result is 
\vspace{1mm}
{\small
\begin{align}
T^{\mu\nu}(x_1) {\cal O}(x_2) &=\Big( A^{\mu\nu} +B^{\mu\nu}(\d) + C^{\mu\nu}(\d^2) +  D^{\mu\nu}(\d^3) + E^{\mu\nu}(\d^4)+  F^{\mu\nu}(\d^5)+ \dots \Big)  {\cal O}(x_2) 
\end{align}
}
\vspace{1mm}
where
\vspace{1mm}
{\small
\begin{align}
A^{\mu\nu}=& {a\over s^4}  \Big(\frac{s^{\mu}s^{\nu}}{s^2}-\frac{\delta^{\mu\nu}}{4} \Big)  \ , \\
B^{\mu\nu} =&  {a \over  \Delta} \Big( {s^\mu \partial^\nu + s^\nu \partial^\mu  \over 2   s^4 }+  {(2 s^\mu s^\nu- s^2 \delta^{\mu\nu})  (s \d) \over 2  s^6}  \Big) \ , \\
\label{Cterm}
C^{\mu\nu} =& {a \over  \Delta (\Delta+1)} \Big( {2\d^\mu \d^\nu+\delta^{\mu\nu}  \Box \over 8 s^2} + {s^\mu  (s \d) \d^\nu +s^\nu  (s \d) \d^\mu\over s^4} - {3  \over 4} {s^\mu s^\nu \Box + \delta^{\mu\nu} (s \d)^2\over s^4} +  {s^\mu s^\nu (s \d)^2 \over s^6} \Big)  \ .
\end{align}\vspace{1mm}
}See Appendix for $D \sim F$ terms. We have denoted
\vspace{1mm}
\begin{align}
(s \d)= s^{\alpha}\p_\alpha , ~~ (s \d)^2=s^{\alpha } s^{\beta } \p_\alpha \p_\beta   , ~~ (s \d)^3 = s^{\alpha } s^{\beta } s^{\gamma } \p_\alpha \p_\beta \p_\gamma  , ~~~ {\rm {etc}. }
\end{align}  

We will find that, in the limiting procedure defined below, only $A \sim C$ terms contribute to the commutators. 
To check this, one needs higher-order terms in the OPE. 
Adopting a different limiting procedure may require higher-order terms. We have listed them (up to the $F$-term) in Appendix for future reference.

To extend the $d=2$ commutator derivation to $d=4$, we first need to define a higher-dimensional analogue of the Virasoro modes. 

\subsubsection*{\it{3.2. Central Terms and Mode Operators on a Plane $\bf \Sigma$}}

In the traditional convention, $z$ and  $\bar z$ are the Euclidean analogue of the lightcone coordinates. 
We will slightly abuse the notation and denote $ds^2= dx^+ dx^- +  \sum_{i=1,2} (d {x_\perp^{(i)}})^2$ in the Euclidean. 
Instead of writing $T^{zz}$ or $T^{\zbar \zbar}$, we will continue to write $T^{\pm \pm}$ in what follows.

Consider the configuration where operators live on a $d=2$ Euclidean plan, ${\bf \Sigma}$, in $d=4$ theories.  Transverse coordinates of the operators are set to zero, i.e. $f(x^+, x^-, x^\perp)|_{\bf \Sigma}\equiv f(x^+, x^-, 0)$. In this type of $d>2$ computation, UV divergences always appear. To motivate our mode operators and limiting procedure, let us first look at central terms. 

The $d=4$ stress-tensor two-point function is given by 
\vspace{1mm}
{\small
\begin{align}
\langle T^{\mu\nu}(x_1) T^{\si\rho}(x_2) \rangle&=C_T {\I^{\mu\nu,\si\rho}(s)\over s^{8}} \ , \\
 {\cal I}^{\mu\nu,\sigma \rho}(s) &= {1\over 2} \Big( { I}^{\mu\sigma}(s) { I}^{\nu\rho}(s)+ { I}^{\mu\rho}(s) { I}^{\nu\sigma}(s)\Big)-{1\over 4} \delta^{\mu\nu} \delta^{\sigma \rho}  \ ,\\
{ I}^{\mu\sigma}(s) &= \delta^{\mu\sigma}- 2 {s^\mu s^\sigma \over s^2}  \ .
\end{align}}\vspace{1mm}We focus on the lightcone components, $T^{++}$ and $T^{--}$.\footnote{We have not exhaustively explored mode operators defined by integrals of $T^{\pm \perp}$, $T^{+-}$, and $T^{\perp\perp}$.  Note that the component $T^{+-}$ can be related to $T^{\perp \perp}$ by the tracelessness condition and thus the situation is more involved. But we expect that these components are irrelevant to the near-lightcone correlator.} On the plane, the stress-tensor correlators have simple forms:
\vspace{1mm}
{\small
\begin{align}
\label{TppTpp}
\langle T^{++}(x_1) T^{++}(x_2) \rangle|_{\bf \Sigma}&={4 C_T \over  (s^-)^6 (s^+)^2}  \ ,\\
\langle T^{++}(x_1) T^{--}(x_2) \rangle|_{\bf \Sigma}&=0  \ ,  \\
\label{TmmTmm}
\langle T^{--}(x_1) T^{--}(x_2) \rangle|_{\bf \Sigma}&={4 C_T \over  (s^-)^2 (s^+)^6}  \ .
\end{align}}\vspace{1mm}We are interested in the lightcone limit: 
\begin{align}
\label{LC}
s^+  = \epsilon \to 0   \ . 
\end{align}
One may interpret this limit as an equal ``time" condition.  The two-point functions \eqref{TppTpp} and \eqref{TmmTmm} on the plane become divergent in this limit. 
To remove the divergence, we consider the following mode operators:  
\vspace{1mm}
{\small
\begin{align}
\label{defK}
{K_m}\Phi (x)|_{\bf \Sigma} & \equiv - {  \pi^2\over 2 } \lim_{x_T^+ \to x^+}  \epsilon   ~ \oint_{{\cal C}(x^-)} {dx_T^-\over 2\pi i} (x_T^-)^{m+2}  ~T^{++} (x_T^{+}, x_T^-, x^\perp_T=0 )
\Phi  (x^+, x^- ,x^\perp=0) \ , \\   
\label{defR}
{R_m} \Phi (x)|_{\bf \Sigma}&  \equiv - { \pi^2 \over 2}  \lim_{x_T^+ \to x^+}  \epsilon^3  ~\oint_{{\cal C}(x^-)} {dx_T^-\over 2\pi i} (x_T^-)^{m}   ~ T^{--} (x_T^{+}, x_T^-, x^\perp_T=0 )
\Phi  (x^+, x^- ,x^\perp=0) \ .
\end{align}}\vspace{1mm}The dimensions of ${K_m}$ and ${R_m}$ are the same. The operator ${K_m}$ formally looks similar to the $d=2$ anti-holomorphic Virasoro operator. But since the lightcone limit \eqref{LC} is imposed on both $T^{++}$ and $T^{--}$, the operator ${R_m}$ does not mimic the $d=2$ holomorphic Virasoro operator.\footnote{Similarly, we adopt $ \lim_{x^+ \to x_T^+}  (x^+ - x_T^+)$ and $ \lim_{x^+ \to x_T^+}  (x^+ - x_T^+)^3$  as the limiting procedures for $\Phi (x){K_m}|_{\bf \Sigma}$ and $\Phi (x){R_m}|_{\bf \Sigma}$, respectively.} An interesting consequence of these definitions is that, as we will see, the $x^+$ dependence automatically disappears from  the $\langle TT \rangle$ and $T{\cal O}$ OPE computations. 

We may proceed to compute the central terms by assuming that the conjugated operators are $K^{\dagger}_m=K_{-m}$ and $R^{\dagger}_m= R_{-m}$.\footnote{One can verify $\langle  K_{m} K_n\rangle|_{\bf \Sigma}\neq 0$ only when $m > 2$ and $n< -2$ while  $\langle  R_{m}  R_n\rangle|_{\bf \Sigma}\neq 0$ only when $m > 0$ and $n< 0$.
For a discussion on the Hermitian conjugate for this type of $d=4$ mode operators, see \cite{Belin:2020lsr}.} 
We find
\begin{align}
\label{centralK}
  \langle [K_m, K_{n}] \rangle|_{\bf \Sigma}&= -  {  \pi^4 \over 120} C_T~  m (m^2-1) (m^2-4) \delta_{m+n,0}  \ ,\\
\label{centralR}
  \langle [{R_m}, R_{n}] \rangle|_{\bf \Sigma}&= - {  \pi^4 }C_T~  m \delta_{m+n,0}  \ . 
\end{align}
As observed in a recent paper \cite{Huang:2021hye}, the $d=4$ structure \eqref{centralK} is identical to the central term of the ${\cal W}_3$ algebra in $d=2$ CFT \cite{Z1985}, and one can exactly reproduce the $d=4$ single stress-tensor exchange contribution via a mode summation using such a central term.  See also \cite{Karlsson:2021mgg} for related observations.  

Let us here make a transformation:
\begin{align}
K_m \to {\widetilde K_m} = {1\over m-2}K_m \ . 
\end{align}This gives 
\begin{align}
  \langle [\widetilde  K_m, \widetilde   K_{n}] \rangle|_{\bf \Sigma}&= { \pi^4 \over 120 } C_T ~  m (m^2-1) \delta_{m+n,0}  \ .
\end{align}
We will show that, by using the $d=4$ $T{\cal O}$ OPE,  the rescaled operator $\widetilde K_m$ and $R_m$ satisfy a Virasoro-Kac-Moody-type algebra. 

\subsubsection*{\it{3.3. Commutators}}

We would like to compute the commutators $[\widetilde K_m,   \widetilde K_n]|_{\bf \Sigma}$, $ [R_m,   R _n]|_{\bf \Sigma}$, and $ [\widetilde K_m,   R_n]|_{\bf \Sigma}$ by applying the $d=4$ $T{\cal O}$ OPE twice.  

We first have
\vspace{1mm}
{\small
\begin{align}
\label{KOOC}
{ \langle{K_m \cal O} (x) {\cal O} (0)\rangle\over \langle  {\cal O} (x) {\cal O} (0)\rangle } \Big|_{\bf \Sigma}
&=    -{ \pi^2\over 2} \oint_{{\cal C}(x^-)} {dx_T^-\over 2\pi i} (x_T^-)^{m+2} \lim_{x_T^{+} \to x^+ }  \epsilon~ {\langle{ T^{++} (x_T^{+}, x_T^-, x^\perp_T=0 )\cal O} (x^{+}, x^-, x^\perp=0) {\cal O} (0)\rangle\over \langle {\cal O} (x^{+}, x^-, x^\perp=0) {\cal O} (0)\rangle} \nn\\
&=  \frac{\Delta}{6}   (m-2) (m-1) (x^{-})^m  + {\cal O} (\epsilon) \ .
\end{align}}\vspace{1mm}Hence 
\vspace{1mm}
{\small
\begin{align}
\label{tildeKOO}
{ \langle{\widetilde K_m \cal O} (x) {\cal O} (0)\rangle\over \langle  {\cal O} (x) {\cal O} (0)\rangle } \Big|_{\bf \Sigma} =  \frac{\Delta}{6}    (m-1)  (x^{-})^m  + {\cal O}(\epsilon) \ .
\end{align}}\vspace{1mm}This result can be obtained using either the exact $\langle T {\cal O} {\cal O} \rangle$ or the $T{\cal O}$ OPE. The operation $\lim_{x_T^{+} \to x^+ }  \epsilon$ is done before performing the contour integral. We will adopt this order of operations.\footnote{In this work, we focus on the simpler case where $x^\perp=0$ for all $d=4$ fields with one scalar sitting at the origin. It would be interesting to identify an algebraic structure in a more general setup.} The factor $(m-1)$ also appears from $\langle L_{m} {\cal O} {\cal O}\rangle$ if one considers the $d=2$ Virasoro mode operator.   

As remarked earlier, our $d=4$ limiting procedure removes the scalar lightcone coordinate $x^+$. This turns out to be a general feature.  
Note that the explicit appearance of the scalar coordinate $x^+$ would be an obstacle to having an algebraic structure: the ``expected" scalings   $\langle \widetilde K_m \widetilde K_n {\cal O} {\cal O}\rangle  \sim  (x^+)^2$ and $\langle \widetilde K_{m+n} {\cal O} {\cal O}\rangle \sim  (x^+)$ in the lightcone limit $x^+ \to 0$ are 
inconsistent with the algebraic structure $\langle [\widetilde K_m, \widetilde K_n ] {\cal O} {\cal O}\rangle  \sim  \langle \widetilde K_{m+n} {\cal O} {\cal O}\rangle$.\footnote{To compute the near-lightcone scalar correlator, one should first note that the scalar's lightcone limit and the $T{\cal O}$ OPE limit generally do not commute; this can be checked using the three-point function $\langle T(x_1) {\cal O}(x_2) {\cal O}(0)\rangle$. Certain scaling rules on mode operators thus might be needed: one may adopt  $\widetilde K_m \to x^+ \widetilde K_m$ to produce the single stress-tensor exchange contribution.  The operator $R_m$ is related to $T^{--}$ and a scaling using a higher power of $x^+$ can suppress its contribution. Because we will not compute stress-tensor exchanges in the present note, these rules will not be discussed.}  

Next, we extend the computation to include two stress tensors. 
Denote\vspace{1mm}
\begin{align}
\Theta^{\mu\nu, \lambda\rho}= { \langle{ T^{\mu\nu} (x_{T1}^{+}, x_{T1}^-, x^\perp_{T1}=0 )T^{\lambda\rho} (x_{T2}^{+}, x_{T2}^-, x^\perp_{T2}=0 )\cal O} (x^{+}, x^-, x^\perp=0) {\cal O} (0)\rangle\over \langle {\cal O} (x^{+}, x^-, x^\perp=0) {\cal O} (0)\rangle}\Big|_{T{\cal O}}
\end{align}\vspace{1mm}
which contains only the $T{\cal O}$ OPE contribution. We obtain\vspace{1mm}
{\small
\begin{align}
\label{KKOO}
&{ \langle  K_m  K_n {\cal O} (x) {\cal O} (0)\rangle\over \langle  {\cal O} (x) {\cal O} (0)\rangle } \Big|_{T{\cal O}, {\bf \Sigma}}\nn\\
&=  { \pi^4\over  4 } \oint_{{\cal C}(x^-)} {dx_{T1}^-\over 2\pi i} (x_{T1}^-)^{m+2}  \oint_{{\cal C}(x^-)} {dx_{T2}^-\over 2\pi i} (x_{T2}^-)^{m+2}  \lim_{x_{T1}^{+} \to x^+ }   \epsilon  \lim_{x_{T2}^{+} \to x^+ } \epsilon ~\Theta^{++,++} \\
&= \frac{\Delta (n-2) (n-1) }{ 36  (\Delta +1)}   \Big(\Delta^2 (m-2) (m-1)+\Delta  (m-2) (6n+m-1)+6 n (m+2 n)\Big)  (x^-)^{m+n} + {\cal O}(\epsilon)  \ .\nn 
\end{align}
}\vspace{1mm}This computation requires the first three terms in the $T{\cal O}$ OPE (i.e. $A, B, C$ terms). In our limiting procedure, higher-order terms do not contribute to \eqref{KKOO}. Since the third-order $C$-term contributes, the result \eqref{KKOO} has a pole at $\Delta= -1$.  This prevents an algebraic structure $[K_m, K_n] \sim K_{m+n}$ for a general $\Delta$. We will take a limit on $\Delta$ to search for emergent algebras. In a more general case, a non-linear algebra might be possible, but this is beyond the scope of this paper. In two-dimensional CFT, only the first two orders in the OPE contribute and  the corresponding result has no pole.

At large $\Delta$, the commutator of the rescaled operator takes a simple form:
\vspace{1mm}
{\small
\begin{align}
\label{kkoo1}
\lim_{\Delta\to \infty}{[ \langle \widetilde K_m,   \widetilde K_n] {\cal O} (x) {\cal O} (0)\rangle\over \langle  {\cal O} (x) {\cal O} (0)\rangle } \Big|_{\bf \Sigma}= {\Delta\over 6}    (m-n) (m+n-1)  (x^-)^{m+n} + {\cal O}(\Delta^0) \ .
\end{align}}\vspace{1mm}An overall sign has been flipped due to the relation \eqref{formula}. The $C$-term in the OPE still contributes at large $\Delta$. 
 The above expression may be identified, if we recall \eqref{tildeKOO}, as
\begin{align}
\label{idK}
(m-n){ \langle{\widetilde K_{m+n} \cal O} (x) {\cal O} (0)\rangle\over \langle  {\cal O} (x) {\cal O} (0)\rangle } \Big|_{\bf \Sigma}  \ .
\end{align}
Including the central term \eqref{centralK} (valid for any $\Delta$), we deduce 
\begin{align}
 [\widetilde K_m,   \widetilde K_n]|_{\bf \Sigma}= (m-n) \widetilde K_{m+n} +  {  \pi^4 \over 120} C_T~  m (m^2-1) \delta_{m+n,0} \ .
\end{align}
In this computation, we set the transverse coordinates  to zero and assume that stress tensors are inserted in a scalar correlator with a large $\Delta$.\footnote{Although the $T{\cal O}$ OPE approach provides hints of having an algebraic structure, there is a potential ambiguity in the identification \eqref{idK} if another operator in the theory also produces the right hand side of \eqref{kkoo1}. It would be useful to sharpen our limiting procedure and isolate the $T^{++}$ contribution from other components since $T^{++}$ is expected to be the dominating contribution to the near-lightcone correlator.}

We next compute
\vspace{1mm}
{\small
\begin{align}
\label{ROOC}
&{ \langle{R_m \cal O} (x) {\cal O} (0)\rangle\over \langle  {\cal O} (x) {\cal O} (0)\rangle } \Big|_{\bf \Sigma}=    {- \pi^2\over 2} \oint_{{\cal C}(x^-)} {dx_T^-\over 2\pi i}  (x_T^-)^{m} \lim_{x_T^{+} \to x^+ }  \epsilon^3~ {\langle{ T^{--} (x_T^{+}, x_T^-, x^\perp_T=0 )\cal O} (x^{+}, x^-, x^\perp=0) {\cal O} (0)\rangle\over \langle {\cal O} (x^{+}, x^-, x^\perp=0) {\cal O} (0)\rangle} \nn\\
&~~~~~~~~~~~~~~~~~~~~~~~ =  \frac{\Delta  }{3}   (x^{+})^m  + {\cal O} (\epsilon)  \ ,\\
\label{RROO}
&{\langle  R_m  R_n {\cal O} (x) {\cal O} (0)\rangle\over \langle  {\cal O} (x) {\cal O} (0)\rangle } \Big|_{TO, {\bf \Sigma}}=  { \pi^4 \over 4} \oint_{{\cal C}(x^-)} {dx_{T1}^-\over 2\pi i}  (x_{T1}^-)^{m}  \oint_{{\cal C}(x^-)} {dx_{T2}^-\over 2\pi i}  (x_{T2}^-)^{m}  \lim_{x_{T1}^{+} \to x^+ }   \epsilon^3  \lim_{x_{T2}^{+} \to x^+ } \epsilon^3 ~\Theta^{--,--} \nn\\
&~~~~~~~~~~~~~~~~~~~~~~~~~~~~~~~ = \frac{\Delta^2  }{  9 }  (x^-)^{m+n} + {\cal O}(\epsilon) \ . 
\end{align}
}\vspace{1mm}These results do not assume a large $\Delta$, and they are fixed by the first term of the $T{\cal O}$ OPE.  Without an operator contribution, the commutator $[R_m, R_n]|_{\bf \Sigma}$ only has the central term which we computed in \eqref{centralR}. 

Let us also consider $[K_m, R_n]|_{\bf \Sigma}$. We have 
\vspace{1mm}
{\small
\begin{align}
\label{KROO}
&{ \langle  K_m  R_n {\cal O} (x) {\cal O} (0)\rangle\over \langle  {\cal O} (x) {\cal O} (0)\rangle } \Big|_{T{\cal O}, {\bf \Sigma}}\nn\\
&=  {\pi^4\over 4 } \oint_{{\cal C}(x^-)} {dx_{T1}^-\over 2\pi i}  (x_{T1}^-)^{m+2}  \oint_{{\cal C}(x^-)} {dx_{T2}^-\over 2\pi i}  (x_{T2}^-)^{n} \lim_{x_{T1}^{+} \to x^+ }   \epsilon  \lim_{x_{T2}^{+} \to x^+ } \epsilon^3 ~ \Theta^{++,--}\\
&=\frac{\Delta }{18 (\Delta +1)} \Big(\Delta ^2 (m-2) (m-1)+\Delta  (m-2) (6 n+m-1)+6 n (m+2 n)\Big)  (x^-)^{m+n} + {\cal O}(\epsilon) \nn
\end{align}
}
\vspace{1mm}
which requires the $C$-term in the OPE. We have further 
\vspace{1mm}
{\small
\begin{align}
\label{RKOO}
{ \langle    R_n  K_m {\cal O} (x) {\cal O} (0)\rangle\over \langle  {\cal O} (x) {\cal O} (0)\rangle } \Big|_{T{\cal O}, {\bf \Sigma}}&=  {\pi^4\over 4 } \oint_{{\cal C}(x^-)} {dx_{T1}^-\over 2\pi i}  (x_{T1}^-)^{n}  \oint_{{\cal C}(x^-)} {dx_{T2}^-\over 2\pi i}  (x_{T2}^-)^{m+2}  \lim_{x_{T1}^{+} \to x^+ }   \epsilon^3  \lim_{x_{T2}^{+} \to x^+ } \epsilon~ \Theta^{--,++} \nn\\
&=  \frac{\Delta^2}{18}  (m-2) (m-1)  (x^-)^{m+n}+ {\cal O}(\epsilon)  \ .
\end{align}}\vspace{1mm}The $C$-term in the OPE affects the result \eqref{RKOO} only through $K_m$; other terms are suppressed in the limit $\epsilon \to 0$.  This is why \eqref{RKOO} has no pole at $\Delta=-1$.  But since  \eqref{KROO} has a pole, let us again take a large $\Delta$. 
The commutator using $\widetilde K_m$ has a simple form:
\vspace{1mm}
{\small
\begin{align}
\lim_{\Delta\to \infty}{ \langle    [ \widetilde K_m,R_ n] {\cal O} (x) {\cal O} (0)\rangle\over \langle  {\cal O} (x) {\cal O} (0)\rangle } \Big|_{{\bf \Sigma}} = -n  {\Delta\over 3}( x^-)^{m+n} + {\cal O}(\Delta^0)
\end{align}}\vspace{1mm}which may be identified, if we recall \eqref{ROOC}, as 
\vspace{1mm}
\begin{align}
- n { \langle{R_{m+n} \cal O} (x) {\cal O} (0)\rangle\over \langle  {\cal O} (x) {\cal O} (0)\rangle } \Big|_{\bf \Sigma} \ .
\end{align}

The above computation thus suggests the following commutators:
\vspace{1mm}
{\small
\begin{align}
\label{a1}
 [\widetilde K_m,   \widetilde K_n]|_{\bf \Sigma}=&~ (m-n) \widetilde K_{m+n} +  {  \pi^4 \over 120} C_T~  m (m^2-1) \delta_{m+n,0} \ , \\
\label{a2}
 [ \widetilde K_m,R_ n]|_{\bf \Sigma}  =&~ - nR_{m+n} \ ,\\
\label{a3}
[{R_m}, R_{n}] |_{\bf \Sigma}=&~ - {  \pi^4 }C_T~  m \delta_{m+n,0}  \ .
\end{align}}\vspace{1mm}This set of commutators has the same form as the $d=2$ Virasoro-Kac-Moody algebra. 
The $d=4$ operator $R_m \sim \int x^{m} T^{--}$ plays a role similar to a $d=2$ conserved $U(1)$ current.  

We emphasize again that these results assume the stress tensors are inserted in a scalar correlator with a large $\Delta$.  
It would be interesting to find an independent derivation of these results and remove potential ambiguities. The mode operators and limiting procedure adopted in this paper are largely guided by having finite central terms. 

In the above computation, we set transverse coordinates to zero after performing the derivatives in the $d=4$ $T {\cal O}$ OPE. 
However, we find that the final results are unchanged if the transverse derivatives and transverse coordinates are set to zero from the start. The $x^\perp$ dependence is subleading in the limit we described. 
This leads to a simplification. We can express $\widetilde K_m$ and $R_m$ as effective differential operators acting on $\Phi(x^+, x^-)=\langle{\cal O} (x^+, x^-) {\cal O} (0)\rangle$:\vspace{1mm}
{\small
\begin{align}
\widetilde K_m \Phi &\simeq
{1\over (m-2)} \Big( 
{\Delta\over 6}  (m+1) (m+2)   (x^-)^{m} + 
(m+2)   (x^-)^{m+1} \d_{x^-} +  \frac{ 2 (x^-)^{m+2}}{(\Delta +1)}\d^2_{x^-}\Big)\Phi  \ , \\  
R_m\Phi &\simeq  {\Delta \over 3} (x^-)^{m} \Phi\ .
\end{align}}\vspace{1mm}They reproduce the non-central terms of \eqref{a1}-\eqref{a3}.  

\subsection*{4. Discussion}

Is there a symmetry underlying the near-lightcone correlator in $d>2$ CFTs with an Einstein gravity dual? The Virasoro symmetry explains why the two-dimensional scalar four-point correlator has the pattern given in \eqref{PANA}. An algebraic approach in $d>2$ will allow us to efficiently compute the multi-stress-tensor contributions to the CFT correlator in higher dimensions, generalizing the notion of $d=2$ Virasoro blocks at large central charge. The structure of multi-stress-tensor operators are intimately related to the stress-tensor commutators. Resolving the tension between divergent central terms in the $d>2$ stress-tensor commutators and the well-behaved scalar correlators presumably requires identifying suitable mode operators and limiting procedure.

This work may be regarded as a small step towards an algebraic derivation of the near-lightcone correlator in higher dimensions. We have shown that mode operators $\widetilde  K_m$ and $R_m$ on a $d=2$ plane satisfy a Virasoro-Kac-Moody-type algebra when the scalar has a large dimension $\Delta$.  It would be interesting to improve our analysis and  incorporate a more general $\Delta$.  One might allow the possibility of having a non-linear algebra.  It is essential to clarify what kinds of mode operators contribute to the near-lightcone correlator.  

Since we focus on a large $\Delta$ limit in this work, it would be interesting to make a connection to the recent discussion on the eigenstate thermalization hypothesis in large $N$ CFTs above two dimensions \cite{Lashkari:2016vgj, Karlsson:2021duj}. In particular, in \cite{Karlsson:2021duj}, it was shown that multi-stress-tensor operators thermalize.  The leading large $\Delta$ structure of the scalar four-point correlator is fixed by the exponentiation of the single stress-tensor exchange, which can be obtained via a mode summation using the $d=4$ central term \eqref{centralK} \cite{Huang:2021hye}.

Our approach relies on the $T{\cal O}$ OPE and we have not utilized the $TT$ OPE in this paper.  The problem is how to compute mode-operator commutators on a $d=2$ plane using the $d=4$ $TT$ OPE or stress-tensor three-point function, $ \langle TTT \rangle$, in the scheme where transverse coordinates are set to zero.  We suspect that it might not be sensible to set all the transverse coordinates to zero in the $TT$ OPE or $ \langle TTT \rangle$ when computing a commutator. In this case, a modified limiting procedure might be required. The stress-tensor three-point function is important because it provides information about the graviton mixing. In $d=2$, the mixing has the form $ \langle L_{m} L_{m} L_{-m-n}\rangle$, which contributes to the two stress-tensor exchanges; see \cite{Fitzpatrick:2015qma, Karlsson:2021mgg} for the $d=2$ discussion. To compute the $d=4$ scalar correlator at subleading $\Delta$, a similar graviton-mixing contribution should be included. 

Another related question concerns the CFT definition of the stress-tensor composite primary operators, $ [T^n]$, for $n\geq 2$ in $d=4$ CFTs.

We do not address these questions here, but we view them as ripe for future study.

\newpage

\subsection*{Acknowledgments} 

I would like to thank L. Fitzpatrick,  G. Korchemsky, A. Parnachev, and A. Zhiboedov for related discussions. 
This work was supported in part by the U.S. Department of Energy Office of Science DE-SC0015845 and the Simons Collaboration on the Nonperturbative Bootstrap. 

\section*{Appendix}

Here we list the $d=4$ $T{\cal O}$ OPE used in Section 3: 
\vspace{1.5mm}
{\small
\begin{align}
T^{\mu\nu}(x_1) {\cal O}(x_2) &=\Big( A^{\mu\nu} +B^{\mu\nu}(\d) + C^{\mu\nu}(\d^2) +  D^{\mu\nu}(\d^3) + E^{\mu\nu}(\d^4)+  F^{\mu\nu}(\d^5)+ \dots \Big)  {\cal O}(x_2) 
\end{align}
\small
}\vspace{1mm}where ($ a= - \frac{2\Delta}{3\pi^2}$)
\vspace{1mm}
{\footnotesize
\begin{align}
A^{\mu\nu}=& {a\over s^4}  \Big(\frac{s^{\mu}s^{\nu}}{s^2}-\frac{\delta^{\mu\nu}}{4} \Big) \ , \\
B^{\mu\nu} =&  {a \over  \Delta} \Big( {s^\mu \partial^\nu + s^\nu \partial^\mu  \over 2   s^4 }+  {(2 s^\mu s^\nu- s^2 \delta^{\mu\nu})  (s \d) \over 2  s^6}  \Big) \ , \\
\label{Cterm}
C^{\mu\nu} =& {a \over  \Delta (\Delta+1)} \Big( {2\d^\mu \d^\nu+\delta^{\mu\nu}  \Box \over 8 s^2} + {s^\mu  (s \d) \d^\nu +s^\nu  (s \d) \d^\mu\over s^4} - {3  \over 4} {s^\mu s^\nu \Box + \delta^{\mu\nu} (s \d)^2\over s^4} +  {s^\mu s^\nu (s \d)^2 \over s^6} \Big) \ , \\
D^{\mu\nu}=&{a \over  \Delta (\Delta+1) (\Delta+2)} 
\Big( {6 (s\d) \d^\mu \d^\nu - 3 ( s^\mu \d^\nu + s^\nu \d^\mu - \delta^{\mu\nu}   (s\d) )\Box\over 8 s^2} \nn\\
&~~~~~~~~~~~~~~~~~~~~~~~~~ + {3\over 2} {s^\mu (s \d)^2 \d^\nu+s^\nu (s \d)^2 \d^\mu -s^\mu s^\nu  (s\d) \Box  \over s^4}  + {(s^\mu s^\nu - \delta^{\mu\nu} s^2) (s \d)^3\over s^6} \Big) \ , \\
 E^{\mu\nu} =&{a \over  \Delta (\Delta+1) (\Delta+2) (\Delta+3)} \Big({-3 \over 16}  \d^\mu \d^\nu \Box
+ {3\over 2 s^2} (s\d)^2 \d^\mu \d^\nu - {9\over 8}{s^\mu   (s\d) \Box\p^\nu + s^\nu   (s\d) \Box \p^\mu   \over s^2}\nn\\
&  ~~~~~~~~~~~~~~~~~~~~~~~~~~~~~~~~~~~+  {2 s^\mu (s\d)^3 \p^\nu + 2s^\nu (s\d)^3 \p^\mu\over s^4} 
+ {3(8 s^\mu s^\nu -  s^2)  \over 64 s^2} \Box^2  \nn\\
&  ~~~~~~~~~~~~~~~~~~~~~~~~~~~~~~~~~~~-{9 s^\mu s^\nu   - 3 s^2 \delta^{\mu\nu}   \over 4 s^4}  (s\d)^2 \Box  + {4 s^\mu s^\nu  -5  s^2 \delta^{\mu\nu}   \over 4  s^6} (s\d)^4
\Big) \ , \\
F^{\mu\nu}=&{a \over  \Delta (\Delta+1) (\Delta+2) (\Delta+3)(\Delta+4)} \Big( {5 s^2 \delta^{\mu\nu}-12 s^\mu s^\nu  \over 4 s^4}(s\d)^3 \Box + {5 (s\d)^3 \p^\mu \p^\nu \over 2 s^2} - {3 (s\d) \Box \d^\mu \d^\nu \over 4}\nn\\
&~~~~~~~~~~~~~~~~~~~~~~~~~~~~~~~~~~~~~~~~~~~~~~ - {3 s^2 \delta^{\mu\nu}-2 s^\mu s^\nu\over 2 s^6} (s\d)^5  - {3\over 16}{  s^2 \delta^{\mu\nu}-6 s^\mu s^\nu  \over s^2}( s\d) \Box^2\nn\\
&~~~~~~~~ + {{3\over 16} ( {s^\mu \Box^2 \d^\nu + s^\nu \Box^2 \d^\mu}) }- {9\over 4} { s^\mu (s\d)^2 \Box \d^\nu + s^\nu (s\d)^2 \Box \d^\mu \over  s^2}+{5\over 2} { s^\mu (s\d)^4 \d^\nu + s^\nu (s\d)^4 \d^\mu\over s^4}\Big) \ .
\end{align}}

\newpage

\bibliographystyle{utphys}
\bibliography{KacTO}

\end{document}